# Broadband ferromagnetic resonance of $Ni_{81}Fe_{19}$ wires using a rectifying effect

A. Yamaguchi [1, 2), *], K. Motoi [1) , *], H. Miyajima [1), *], Y. Miyashita [1)] and Y. Sanada [3)]

[1)] Department of Physics, Keio University, Hiyoshi 3-14-1, Yokohama 223-8522, Japan

[2)] PRESTO, JST, Sanbancho 5, Chiyoda, Tokyo 102-0075, Japan

[3)] Department of Electronics and Electrical Engineering, Keio University, Hiyoshi 3-14-1, Yokohama 223-8522, Japan

* Corresponding authors' E-mail: yamaguch@phys.keio.ac.jp (A. Y.), kmotoi@phys.keio.ac.jp (K.M.), and miyajima@phys.keio.ac.jp (M. H.)

**[Abstract]**

The broadband ferromagnetic resonance measurement using the rectifying effect of $Ni_{81}Fe_{19}$ wire has been investigated. One wire is deposited on the center strip line of the coplanar waveguide (CPW) and the other one deposited between two strip lines of CPW. The method is based on the detection of the magnetoresistance oscillation due to the magnetization dynamics induced by the radio frequency field. The magnetic field dependences of the resonance frequency and the rectification spectrum are presented and analytically interpreted on the standpoint of a uniform magnetization precession model.

## I. Introduction

Ferromagnetic properties in the GHz region have been actively explored for applications in radio-frequency (RF)/microwave devices by using many kinds of procedure, such as ferromagnetic resonance (FMR), Brillouin light scattering [1 – 3], the time-resolved magneto-optical Kerr effect [4 - 6], and so on. These procedures, however, have their own distinguishing advantages and disadvantages for the characterization of the high-frequency magnetization dynamics. The development of new methods as well as measuring techniques is a crucial issue, not only for the basic scientific field but also technical applications.

Recently, several new electrical measurements [7 – 13] were proposed, which are extremely sensitive and suitable for investigating the magnetization dynamics in sub-micron scale magnets. One of them is pulse inductive measuring used to determine the time domain [9] of dynamical properties of component films and multilayer spin-valve stacks. It is especially useful in the characterization of the intrinsic dynamical properties with recourse to expensive storage-oscilloscopes for high-speed sampling and wideband amplifiers. The broadband spectrometer is preferable since it allows one to study FMR in constant magnetic fields. The novel inductive technique of FMR using vector network analyzer gives an insight into the modal spectrum with respect to the frequency and effective damping of the various modes [7, 8].

Currently, the investigation of spin-polarized current has progressed, with applications to spintronics devices such as magneto-resistive random-access memories and microwave generators. The spin-polarized current flowing through ferromagnetic multilayers is known to generate spin wave excitation [14, 15] together with magnetization reversal. The physical mechanism, as revealed by the experimental results, is a consequence of the spin angular momentum transfer, which occurs due to the interaction between the spin-polarized current and the ferromagnetic moment. One of the interesting properties is the rectification of RF current occurring in magnetic tunnel junction (MTJ) [10], in spin-valve structure [11] and in single-layered ferromagnetic wire [12]. This rectification is explained in terms of the by the magnetoresistance oscillation attributable to spin-transfer torque. It should be noted that direct-current (DC) voltage is produced whenever the resonant RF current flows through these systems.

In this study, we present the broadband FMR induced by RF current flowing in the coplanar waveguide (CPW) and show the experimental results on sub-micron single-layered ferromagnetic $Ni_{81}Fe_{19}$ wires. We also discuss the FMR spectrum, excited not only by the magnetic field but also by the difference in spin-transfer torque. In order to investigate the spin dynamics induced by the in-plane and out-of-plane RF field components, two kinds of devices were prepared; one is $Ni_{81}Fe_{19}$ wire, prepared on the center strip line of the CPW, and the other is prepared between the center strip line and the ground strip line of the CPW.

## II. Sample fabrication and magnetic field mode

The 22 and 30-nm-thick polycrystalline $Ni_{81}Fe_{19}$ wires are patterned by electron beam lithography and lift-off processing (Fig. 1). Two kinds of wire are prepared; One is prepared on the top of the CPW comprising a Cr (5 nm)/Au (38 nm) conductive strip (Fig. 1(a)), and the other is prepared in the aperture between the conductive strip lines of the CPW comprising Cr (5 nm)/Au (80 nm) on MgO (100) substrates (Fig. 1(b)). The width of the 22nm-thick $Ni_{81}Fe_{19}$ wire prepared on the CPW (Fig. 1(a)) is 2 μm, and that of the CPW strip lines is 3 μm. On the other hand, the width of the 30nm-thick wire prepared between the strip lines (Fig. 1(b)) is 2 μm, and that of the CPW line is 5 μm.

The ground-signal-ground (GSG) type microwave probe is connected to the CPW, and the DC voltage difference induced by the RF current flowing through the system [12] is detected by using a bias-tee circuit and a voltmeter, as illustrated in Fig. 1(c). The sinusoidal voltage output of the signal generator (SG), with frequency range from 10 MHz to 15 GHz, produces an approximately elliptical field pattern $h_{\mathrm{RF}}$ around the central line. In the arrangement in Fig. 1(a), the RF current flowing through the center strip gives rise to the transverse RF field, while in Fig. 1(b), the RF field acting on the wire is perpendicular to the plane due to the application of the RF current across the CPW. The external static magnetic field $\boldsymbol{H}_{\mathrm{ext}}$ is applied in the substrate plane as a function of tilting angle $\theta$ from the longitudinal axis of the wire and the measurements are performed at room temperature.

Figures 2(a) and 2(b) show the RF field distributions obtained by the High-Frequency Structure Simulator (HFSS) [17] for the cases of Figs. 1(a) and 1(b), respectively. The geometry and dimensions of the calculation are the same as those of the present sample except for the magnitude of the applied microwave power. The RF field distribution whithin the real space, visually shown in Figs. 2(a) and 2(b), is consistent with the analytical model and the analysis for the experimental results as latterly described.

As shown in Fig. 2(a), the in-plane Oersted field along the $y$ direction, produced by the RF current $I$ flowing at a distance $z$ is $H_y$ $(I, z)$ in the configuration of Fig. 1(a), where the $y$ axis is transverse to the direction of the current. At a position close to the center conductor of the CPW but far from an edge, the center conductor appears as an infinite sheet of current, which produces the field $h_{xy} = I/2w$ [9], where $w$ is the width of the center conductor of the CPW. The arrangement of Fig. 1(b) also provides an out-of-plane field $H_z(I, y) = h_z = I/2y$ along the wire axis, as shown in Fig. 2(b), in which $y$ is the distance between the center of the wire and the CPW center strip line.

## Ⅲ Analytical model for the spin dynamics

The dynamics of the magnetic moment in RF field are analytically described by the Landau-Lifshits-Gilbert (LLG) equation. The LLG equation in the coordinate system shown in Fig. 1(e) is expressed by

$$\frac{\partial \boldsymbol{m}(\mathrm{t})}{\partial \mathrm{t}} = -\gamma_0 \boldsymbol{m}(\mathrm{t}) \times \left(\boldsymbol{H}_{\mathrm{eff}} + \boldsymbol{h}_{\mathrm{RF}}\right) + \alpha \boldsymbol{m}(\mathrm{t}) \times \frac{\partial \boldsymbol{m}(\mathrm{t})}{\partial \mathrm{t}} , \tag{1}$$

where $\boldsymbol{m}(t)$ denotes the unit vector along the local magnetization, ($\boldsymbol{m} = \boldsymbol{M}/M_{\mathrm{S}}$ and $|\boldsymbol{m}| = 1$), $\gamma_0$ the gyromagnetic ratio, $\boldsymbol{H}_{\mathrm{eff}}$ the effective magnetic field, including the exchange and demagnetizing fields, $\boldsymbol{h}_{\mathrm{RF}}$ the RF field produced by the RF current flowing through the center strip of the CPW, and $\alpha$ the Gilbert damping constant.

As schematically shown in Fig. 1(e), we define the $(x, y, z)$ coordinate system, in which each component corresponds to the vertical wire axis, the longitudinal wire axis, and the normal to the substrate plane, respectively. The external magnetic field is directed at angle $\theta$ from the $x$-coordinate axis. Subsequenctly, we also define the coordinate system $(a, b, c)$ where the $a$ direction corresponds to the equilibrium direction of $\boldsymbol{m}_0$ along the effective magnetic field $\boldsymbol{H}_{\mathrm{eff}} = \boldsymbol{H}_{\mathrm{ext}} + \boldsymbol{H}_{\mathrm{A}}$ including the external field $\boldsymbol{H}_{\mathrm{ext}}$ and the shape magnetic anisotropy field $\boldsymbol{H}_{\mathrm{A}}$. Now, we assume the uniform precession to be dominant and the exchange field to have disappeared.

The unit vector $\boldsymbol{m}_0$ inclines at an angle $\psi$ from the $x$-coordinate axis. The magnetization precession around $\boldsymbol{H}_{\mathrm{eff}}$ results in a small time-dependent component of the magnetization perpendicular to $\boldsymbol{m}_0$, which is almost parallel to the direction of $\boldsymbol{H}_{\mathrm{eff}}$.

Subsequently, we can decompose the unit vector $\boldsymbol{m}(t)$ as

$$\boldsymbol{m}(t)=\boldsymbol{m}_0+\delta\boldsymbol{m}(t)=\left(m_{\mathrm{a}}(t),m_{\mathrm{b}}(t),m_{\mathrm{c}}(t)\right)\ , \tag{2}$$

where $\left|\boldsymbol{m}(t)\right|=1$ and so $m_{\mathrm{a}}=\sqrt{1-\left(m_{\mathrm{b}}^2+m_{\mathrm{c}}^2\right)}$ . The external field $\boldsymbol{H}_{\mathrm{ext}}$, the anisotropy field $\boldsymbol{H}_{\mathrm{A}}$, and the effective field $\boldsymbol{H}_{\mathrm{eff}}$ in the $(a,b,c)$ coordinate system are given by

$$\boldsymbol{H}_{\mathrm{ext}}=\begin{pmatrix} H_{\mathrm{ext}}\cos(\theta-\psi) \\ H_{\mathrm{ext}}\sin(\theta-\psi) \\ 0 \end{pmatrix}, \tag{3}$$

$$\boldsymbol{H}_{\mathrm{A}}=-M_{\mathrm{S}}\tilde{\boldsymbol{N}}\cdot\boldsymbol{m}=-M_{\mathrm{S}}\begin{pmatrix} N_{\mathrm{a}}m_{\mathrm{a}} \\ N_{\mathrm{b}}m_{\mathrm{b}} \\ N_{\mathrm{c}}m_{\mathrm{c}} \end{pmatrix}, \tag{4}$$

and

$$\boldsymbol{H}_{\mathrm{eff}}=\boldsymbol{H}_{\mathrm{ext}}+\boldsymbol{H}_{\mathrm{A}}=\begin{pmatrix} H_{\mathrm{a}} \\ H_{\mathrm{b}} \\ H_{\mathrm{c}} \end{pmatrix}=\begin{pmatrix} H_{\mathrm{ext}}\cos(\theta-\psi)-M_{\mathrm{S}}N_{\mathrm{a}}m_{\mathrm{a}} \\ H_{\mathrm{ext}}\sin(\theta-\psi)-M_{\mathrm{S}}N_{\mathrm{b}}m_{\mathrm{b}} \\ -M_{\mathrm{S}}N_{\mathrm{c}}m_{\mathrm{c}} \end{pmatrix}, \tag{5}$$

where $\tilde{\boldsymbol{N}}$ is the demagnetization coefficient in the $(a,b,c)$ coordinate system, given by

$$\tilde{\boldsymbol{N}}=\begin{pmatrix} N_{\mathrm{a}} \\ N_{\mathrm{b}} \\ N_{\mathrm{c}} \end{pmatrix}=\begin{pmatrix} \cos\psi & -\sin\psi & 0 \\ \sin\psi & \cos\psi & 0 \\ 0 & 0 & 1 \end{pmatrix}\begin{pmatrix} N_{\mathrm{x}} \\ N_{\mathrm{y}} \\ N_{\mathrm{z}} \end{pmatrix}=\begin{pmatrix} N_{\mathrm{x}}\cos\psi-N_{\mathrm{y}}\sin\psi \\ N_{\mathrm{x}}\sin\psi+N_{\mathrm{y}}\cos\psi \\ N_{\mathrm{z}} \end{pmatrix}. \tag{6}$$

In the case of the wire prepared on the conductor strip line, the RF field $\boldsymbol{h}_{\mathrm{RF}\|}$ in the plane is

$$\boldsymbol{h}_{\mathrm{RF}\|}=\left(h_{\mathrm{xy}}\mathrm{e}^{i\omega t}\sin\psi,h_{\mathrm{xy}}\mathrm{e}^{i\omega t}\cos\psi,0\right). \tag{7}$$

While, in the case of the wires prepared between the strip lines, the RF field perpendicular to the plane is written by

$$\boldsymbol{h}_{\mathrm{RF}\perp} = \left(0, 0, h_{\mathrm{z}} \mathrm{e}^{i\omega t}\right). \tag{8}$$

Here, the aspect ratio of the wire elongated along the $x$ axis is large, and consequently $N_{\mathrm{x}}$ is almost zero. In the small precession angle limit around the effective field $\boldsymbol{H}_{\mathrm{eff}}$ ($\mathrm{d}m_{\mathrm{a}}/\mathrm{d}t = 0$, $m_{\mathrm{a}} \simeq 1$), the LLG equation can be linearized. In response to each driving field $\boldsymbol{h}_{\mathrm{RF}\parallel}$ and $\boldsymbol{h}_{\mathrm{RF}\perp}$ with angular frequency $\omega$, we can solve Eq. (1), considering a small variation,

$$\delta\boldsymbol{m} \approx \left(0, m_{\mathrm{b}}(t), m_{\mathrm{c}}(t)\right) \approx \left(0, m_{\mathrm{b}} \mathrm{e}^{i\omega t}, m_{\mathrm{c}} \mathrm{e}^{i\omega t}\right) \text{ and } \boldsymbol{m}_0 \approx (1, 0, 0). \tag{9}$$

The linearized LLG equation for the magnetization is reduced to a partial differential equation for the dynamic magnetization $m_{\mathrm{b}}(t) = m_{\mathrm{b}}\,\mathrm{e}^{i\omega t}$ and $m_{\mathrm{c}}(t) = m_{\mathrm{c}}\,\mathrm{e}^{i\omega t}$. Inserting them into Eq. (1) and taking only the linear terms in $m_{\mathrm{b}}$ and $m_{\mathrm{c}}$ into account, one obtains the following two cases:

**A. The driving field $\boldsymbol{h}_{\mathrm{RF}\parallel}$ in the plane of the wire**

After the substitution of Eqs. (5), (7), and (9) into Eq. (1) and the rearrangement of the terms, the following equation is obtained, neglecting the very small quadratic terms $h_{\mathrm{xy}} m_{\mathrm{c}}$ and $h_{\mathrm{xy}} m_{\mathrm{b}}$:

$$\frac{\partial}{\partial t}\begin{pmatrix} m_{\mathrm{b}}(t) \\ m_{\mathrm{c}}(t) \end{pmatrix} \approx -\gamma_0 \begin{pmatrix} m_{\mathrm{c}}(t)\left(H_{\mathrm{a}} + M_{\mathrm{S}} N_{\mathrm{c}}\right) \\ H_{\mathrm{ext}} \sin(\theta - \psi) - [H_{\mathrm{a}} + M_{\mathrm{S}} N_{\mathrm{b}}] m_{\mathrm{b}}(t) + h_{\mathrm{xy}} e^{i\omega t} \cos\psi \end{pmatrix} + \alpha \frac{\partial}{\partial t}\begin{pmatrix} -m_{\mathrm{c}}(t) \\ m_{\mathrm{b}}(t) \end{pmatrix}. \tag{10}$$

Rearranging Eq. (10), we obtain

$$\begin{cases} i\omega m_{\mathrm{b}} + [\gamma_0 H'_{\mathrm{c}} + i\omega\alpha] m_{\mathrm{c}} = 0 \\ -[\gamma_0 H'_{\mathrm{b}} + i\omega\alpha] m_{\mathrm{b}} + i\omega m_{\mathrm{c}} = -\gamma_0 \left[ H_{\mathrm{ext}} \sin(\theta-\psi) \mathrm{e}^{-i\omega t} + h_{\mathrm{xy}} \cos\psi \right] \end{cases}, \tag{11}$$

where the components of the static field are replaced by

$$\begin{cases} H'_{\mathrm{b}} = H_{\mathrm{a}} + M_{\mathrm{S}} N_{\mathrm{b}} \\ H'_{\mathrm{c}} = H_{\mathrm{a}} + M_{\mathrm{S}} N_{\mathrm{c}} \end{cases}. \tag{12}$$

Forming a queue form Eq.(11), we obtain

$$\begin{pmatrix} i\omega & [\gamma_0 H'_{\mathrm{c}} + i\omega\alpha] \\ -[\gamma_0 H'_{\mathrm{b}} + i\omega\alpha] & i\omega \end{pmatrix} \begin{pmatrix} m_{\mathrm{b}} \\ m_{\mathrm{c}} \end{pmatrix} = \begin{pmatrix} 0 \\ -\gamma_0 \left[ H_{\mathrm{ext}} \sin(\theta-\psi) \mathrm{e}^{-i\omega t} + h_{\mathrm{xy}} \cos\psi \right] \end{pmatrix}, \tag{13}$$

In general, in ferromagnetic conductor is very small, and then $\alpha^2 \approx 0$ in the calculation. The solution of Eq. (13) is given by

$$\begin{pmatrix} m_{\mathrm{b}} \\ m_{\mathrm{c}} \end{pmatrix} \simeq \frac{\gamma_0 \left[ H_{\mathrm{ext}} \sin(\theta-\psi) \mathrm{e}^{-i\omega t} + h_{\mathrm{xy}} \cos\psi \right]}{\left( \omega_{\mathrm{k}}^2 - \omega^2 \right)^2 + \omega^2 \alpha^2 \Delta^2} \begin{pmatrix} \gamma_0 H'_{\mathrm{c}} \left( \omega_{\mathrm{k}}^2 - \omega^2 \right) + i\omega\alpha \left[ \left( \omega_{\mathrm{k}}^2 - \omega^2 \right) - \gamma_0 \Delta H'_{\mathrm{c}} \right] \\ -\omega^2 \alpha \Delta - i\omega \left( \omega_{\mathrm{k}}^2 - \omega^2 \right) \end{pmatrix}, \tag{14}$$

where the FMR frequency $\omega_{\mathrm{k}}$ and $\Delta$ are given by the following relations:

$$\omega_{\mathrm{k}}^2 = \gamma_0^2 H'_{\mathrm{c}} H'_{\mathrm{b}}, \tag{15}$$

$$\Delta = \gamma_0 \left( H'_{\mathrm{b}} + H'_{\mathrm{c}} \right). \tag{16}$$

Now, the RF field is given by $h(t) = I(t)/2w$ with $I(t) = I \cos\omega t$. The dynamic component $\delta m_{\mathrm{b}}$ induces the AMR with a magnetization precession angle of $\phi(t)$ around $\boldsymbol{H}_{\mathrm{eff}}$, namely,

$$\sin\phi(t) \approx \left[ |\delta \boldsymbol{m}| / |\boldsymbol{m}| \right]_{\mathrm{b}} \quad \text{and} \quad \sin 2\phi(t) \approx 2\left[ |\delta \boldsymbol{m}| / |\boldsymbol{m}| \right]_{\mathrm{b}}, \tag{17}$$

and the inductive voltage is $V \propto \mathrm{d}\delta m_{\mathrm{b}}/\mathrm{d}t$. In the present case, a microwave current is flowing, induced in the wire located between the strip lines. The detection circuit has capacitive and inductive coupling to the CPW structure (Fig. 2), as well as the source of rectification between the AMR and induced microwave currents. By considering these conditions, the product of the AMR and the current yields the DC voltage $V_0(t)$.

$$\begin{aligned} V_0(t) &= I(t)\cdot R(t) \\ &= I\cos\omega t\cdot\Delta R\cos^2\left(\psi+\phi(t)\right) \\ &= I\cdot\Delta R\cos\omega t\left[\cos^2\psi+\left(1-2\cos^2\psi\right)\sin^2\phi(t)-\frac{1}{2}\sin 2\psi\sin 2\phi(t)\right] \end{aligned}. \quad (18)$$

The rectifying frequency spectrum of $V_0(t)$ is evaluated by the Fourier transformation of Eq. (18);

$$\begin{aligned} V_0(\omega) &\approx \Delta R\cdot I_{\mathrm{S}}\cdot\left\{-\frac{1}{2}\sin 2\psi\sin 2\phi(t)+\left(1-2\cos^2\psi\right)\sin^2\phi(t)\right\} \\ &\approx \Delta R\cdot I_{\mathrm{S}}\cdot\left\{-\frac{1}{2}\sin 2\psi\cdot 2\left[\frac{|\delta\boldsymbol{m}|}{|\boldsymbol{m}|}\right]_{\mathrm{b}}+\left(1-2\cos^2\psi\right)\left[\frac{|\delta\boldsymbol{m}|}{|\boldsymbol{m}|}\right]_{\mathrm{b}}^2\right\} \end{aligned}, \quad (19)$$

where $I_{\mathrm{S}}$ and $I_{\mathrm{C}}$ are the current flowing through the wire and that flowing through the conductive strip line of the CPW, respectively. The second harmonic term cannot produce rectification and the DC voltage spectrum $V_0(\omega)$ is expressed as

$$V_0(\omega)\approx A(\omega)\cdot\frac{\Delta R\cdot I_{\mathrm{S}}\cdot I_{\mathrm{C}}}{2w}\left[-\frac{1}{2}\sin 2\psi\cos\psi+2\left(1-2\cos^2\psi\right)\cos\psi\cdot A(\omega)\cdot H_{\mathrm{ext}}\sin(\theta-\psi)\right], \quad (20)$$

where $A(\omega)$ is given by

$$A(\omega)\approx\frac{\gamma_0^2 H_{\mathrm{c}}'\left(\omega_{\mathrm{k}}^2-\omega^2\right)}{\left(\omega_{\mathrm{k}}^2-\omega^2\right)^2+\omega^2\alpha^2\Delta^2}. \quad (21)$$

When the external field $|\boldsymbol{H}_{\mathrm{ext}}|$ exceeds the anisotropy field $|\boldsymbol{H}_{\mathrm{A}}|$, the magnetization aligns almost parallel to $\boldsymbol{H}_{\mathrm{ext}}$, namely, $\psi\approx\theta$. Then, $V_0(\omega)$ is proportional

to $\sin 2\theta \cos\theta$. On the other hand, when $|\boldsymbol{H}_{\text{ext}}|$ is smaller than $|\boldsymbol{H}_{\text{A}}|$, the magnetic moment directs along the longitudinal wire axis, namely, $\psi \approx 0^{\circ}$, and $V_0(t)$ is almost proportional to $H_{\text{ext}} \sin\theta$. Consequently, the field and angle dependences of the induced DC voltage $V(\omega)$ are summarized as follows:

(i) $V_0(\omega) \propto \sin 2\theta \cos\theta$ when $|\boldsymbol{H}_{\text{ext}}| >> |\boldsymbol{H}_{\text{A}}|$ (22)

(ii) $V_0(\omega) \propto H_{\text{ext}} \sin\theta$ when $|\boldsymbol{H}_{\text{ext}}| << |\boldsymbol{H}_{\text{A}}|$ (23)

These results are consistent with the previous works. Here, we must consider the effect of the inductive voltage signal (IVS), which is closely related to the transverse magnetization through Faraday's law. The IVS is proportional to $\mathrm{d}m_{\mathrm{y}}/\mathrm{d}t = \cos\psi \cdot \mathrm{d}m_{\mathrm{b}}/\mathrm{d}t$ [9]. In this case, the $\theta$ dependence of the IVS is given by $\cos^2\theta$ and $\sin\theta$ for $|\boldsymbol{H}_{\text{ext}}| >> |\boldsymbol{H}_{\text{A}}|$ and $|\boldsymbol{H}_{\text{ext}}| << |\boldsymbol{H}_{\text{A}}|$, respectively. Therefore, $V_0$ is produced by the AMR oscillation due to the RF field in the former case (i). The IVS, however, may contribute to the DC signal in the latter case (ii). It is important to note the type of $V_0(\omega)$ due to the RF field always takes the dispersion one, while $V_0(\omega)$ caused by the RF spin-polarized current directly flowing through the wire is given by the superposition of the Lorentzian and dispersion types [12].

## B. The driving field $\boldsymbol{h}_{\mathrm{RF}\perp}$ perpendicular to the plane of the wire

Where the wire is located between the center strip line and the ground strip line of the CPW, the RF field $\boldsymbol{h}_{\mathrm{RF}}$ given by Eq. (8) is perpendicular to the plane on the wire position.

Likewise, after the substitution of Eqs. (5), (8), and (9) into Eq. (1) and the rearrangement of the terms we obtain, neglecting the very small quadratic terms $h_z m_c$ and $h_z m_b$,

$$\frac{\partial}{\partial t}\begin{pmatrix} m_b(t) \\ m_c(t) \end{pmatrix} \approx -\gamma_0 \begin{pmatrix} m_c(t) H'_c - h_z e^{i\omega t} \\ H_{ext}\sin(\theta-\psi) - m_b(t) H'_b \end{pmatrix} + \alpha \frac{\partial}{\partial t}\begin{pmatrix} -m_c(t) \\ m_b(t) \end{pmatrix}, \tag{24}$$

and rearranging, we obtain

$$\begin{pmatrix} i\omega & \gamma_0 H'_c + i\omega\alpha \\ -(\gamma_0 H'_b + i\omega\alpha) & i\omega \end{pmatrix}\begin{pmatrix} m_b \\ m_c \end{pmatrix} \approx -\gamma_0 \begin{pmatrix} -h_z \\ H_{ext}\sin(\theta-\psi) e^{-i\omega t} \end{pmatrix}. \tag{25}$$

The real part of the solution of Eq. (25) can be presented by

$$\begin{pmatrix} m_b \\ m_c \end{pmatrix} \approx \frac{\gamma_0\left\{(\omega_k^2-\omega^2) - i\omega\alpha\Delta\right\}}{\left(\omega_k^2-\omega^2\right)^2 + \omega^2\alpha^2\Delta^2} \times \begin{pmatrix} i\omega h_z + (\gamma_0 H'_c + i\omega\alpha) H_{ext}\sin(\theta-\psi) e^{-i\omega t} \\ (\gamma_0 H'_b + i\omega\alpha) h_z - i\omega H_{ext}\sin(\theta-\psi) e^{-i\omega t} \end{pmatrix}. \tag{26}$$

There is the source of rectification between the AMR and induced microwave currents. Multiplying them yields a DC voltage. The frequency spectrum of the induced voltage $V_0(\omega)$ is also expressed by the Fourier transformation of Eq. (26);

$$\begin{aligned} V_0(\omega) &\approx \Delta R \cdot I_S \cdot \left\{ -\frac{1}{2}\sin 2\psi \sin 2\phi(t) + \left(1 - 2\cos^2\psi\right)\sin^2\phi(t) \right\} \\ &\approx \Delta R \cdot I_S \cdot \left\{ -\frac{1}{2}\sin 2\psi \cdot 2\left[\frac{|\delta \boldsymbol{m}|}{|\boldsymbol{m}|}\right]_b + \left(1 - 2\cos^2\psi\right)\left[\frac{|\delta \boldsymbol{m}|}{|\boldsymbol{m}|}\right]_b^2 \right\} \\ &\approx B(\omega)\cdot\omega^2\alpha\cdot\Delta R \cdot \frac{I_C \cdot I_S}{2y} \\ &\quad \times\left\{ -\Delta \sin 2\psi + 2\left(1-\cos^2\psi\right)\cdot B(\omega)\cdot\left(\omega_k^2-\omega^2\right)\cdot\left\{2\Delta\gamma_0 H'_c - \left(\omega_k^2-\omega^2\right)\right\} H_{ext}\sin(\theta-\psi) \right\} \end{aligned} \tag{27}$$

where $I_S$ and $I_C$ are the current flowing through the wire and that flowing through the conductive strip line of the CPW, respectively, and $B(\omega)$ is given by

$$B(\omega) = \frac{\gamma_0}{\left(\omega_k^2-\omega^2\right)^2 + \omega^2\alpha^2\Delta^2}. \tag{28}$$

When $|\boldsymbol{H}_{\text{ext}}|$ is much larger than $|\boldsymbol{H}_{\text{A}}|$, the magnetization directs along the applied field, then $\psi \approx \theta$ and the rectifying voltage can be expressed as

$$V_0(\omega) = -B(\omega) \cdot \frac{I_{\text{S}} \cdot I_{\text{C}} \cdot \Delta R \cdot \omega^2 \cdot \alpha \cdot \Delta}{2y} \cdot \sin 2\theta \, . \tag{29}$$

This condition shows that $V_0(\omega)$ is proportional to $\sin 2\theta$, which is consistent with the result of Costache [13]. Conversely, $|\boldsymbol{H}_{\text{ext}}|$ is smaller than $|\boldsymbol{H}_{\text{A}}|$, the magnetic moment is almost parallel to the direction of $\boldsymbol{H}_{\text{eff}}$. Then, $0^\circ < \psi = \text{const.} < \theta$, and the $V_0(\omega)$ is almost proportional to $\sin\theta$.

## III. Experimental Results

### A. DC voltage induced by the in-plane RF field

In the rectification of RF current in the wire, the magnetization dynamics induced by the RF spin-polarized current result in the resistance oscillation that originated from the AMR effect [12], which generates the DC voltage $V_0$ combining with the RF current. Conversely, in the system comprising wires prepared on the CPW (Fig. 1(a)), the majority of the RF current flows in the CPW line and the wire. As shown in Fig. 1(d), the total resistance $R$ of the system can be calculated by

$$R(t) = R_{\text{S}}(t) R_{\text{C}} / \left(R_{\text{S}}(t) + R_{\text{C}}\right), \tag{30}$$

where $R_{\text{S}}(t)$ and $R_{\text{C}}$ are the resistance of the $Ni_{81}Fe_{19}$ wire and that of the CPW line, respectively. In the system, $V_0$ is given by the multiple of $I(t)$ and $R(t)$. Here, as the resistance of the CPW line is almost constant, $V_0$ is reduced to

$$V_0(t) = I(t)\cdot R(t) \approx I_S(t)\cdot R_S(t), \tag{31}$$

where $I = I_C + I_S$, $I_C = IR_S/(R_C + R_S)$ and $I_S = IR_C/(R_C + R_S)$.

In the present frequency region, the CPW magnetic field distribution is similar to that of the quasi-TEM mode [16] as shown in Figs. 2(a) and 2(b), with the microwave magnetic-field pattern forming an ellipse about the central strip conductor. The microwave power injected into the CPW is +5dBm. The DC resistance of the CPW line is $R_C$ = 55 Ω while that of $Ni_{81}Fe_{19}$ wire is estimated to be $R_S$ = 600 Ω, assuming the resistivity of $Ni_{81}Fe_{19}$ wire to be equal to that of bulk $Ni_{81}Fe_{19}$. Due to the impedance mismatch among cables, the bias tee, the GSG typed microwave probe, and the CPW with the sample, the RF attenuation depends on the frequency. To estimate the losses due to the impedance mismatch, we measure the total impedance of the system as a function of frequency using a vector network analyzer. For a wire of with 2μm prepared on the CPW, the impedance increases with frequency, while the change in impedance can cause an increase of up to 10 percent increase, virtually independently of frequency within the operational frequency range below 15 GHz (see Fig. 3). Therefore, we estimate that each RF current density flowing through the center coplanar strip line and the wire as $(6.0\sim 6.2)\times 10^{10}$ A/m$^2$ and $(7.0\sim 7.2)\times 10^{9}$ A/m$^2$, respectively, the value of which are relatively low for the RF Joule heating at input power of +5dBm within the range 0 to 15 GHz,

The amplitude of the RF field on the center strip line is estimated to be 15 Oe [7-9].

The magnetic field is applied to the substrate plane every $5^{\circ}$ from the wire-axis. Figure 4(a) shows the field dependence of the DC voltage difference $\Delta V_0(\omega)$ ($\theta = 30^{\circ}$) generated in the 2μm width wire prepared on the CPW, which is shown in Fig. 4(a). As seen in the figure, the peak (dip) position of the resonance spectrum shifts toward a higher frequency region with an increasing magnetic field. Figure 4(b) shows the frequency variations of the center between the peak and the dip for different frequencies of the RF field. The (red) solid circle is the experimental data, while the solid line fits with the Kittel formula for the uniform precession mode, including the shape magnetic anisotropy [18];

$$\omega(H) = \frac{g\mu_B\mu_0}{\hbar} \cdot \left[ (H + H_A) \cdot (H + H_A + M_S) \right]^{\frac{1}{2}}, \qquad (32)$$

where $g$ is the Lande factor, and $H_A$ the effective anisotropy field, including demagnetizing and exchange fields. The magnetic field dependence of the resonance frequency is in positive agreement with that of calculation using the Kittel's Eq. (32), which is consistent with Eq. (15). As the RF field induced by the current precesses the magnetic moment in the wire, the AMR resistance also oscillates at the resonance frequency. This resistance oscillation generates $V_0$ combining with the RF current.

The rectifying spectra for the magnetic field of 900 Oe in the substrate plane at $\theta = 0^{\circ}, 35^{\circ}, 85^{\circ}, 95^{\circ}, 145^{\circ}$, and $180^{\circ}$ are shown in Fig. 4(c). As seen in the figure, the spectrum strongly depends on $\theta$, indicating that the uniform precession is affected by $\theta$ and that the sign of $V_0$ reverses when the sense of the field is reversed ($\theta$ to $\theta+180^\circ$).

Figure 5 shows the $\theta$ dependence of $\Delta V_0$ in the static magnetic fields of (a) $|\mathbf{H}_{\text{ext}}|=900$ Oe and (b) $|\mathbf{H}_{\text{ext}}|=30$ Oe. The solid curve represents (a) $\sin 2\theta \cos\theta$ and (b) $\sin\theta$ fitting curves which are obtained from the analytical calculation result of case (i) and (ii) of Eq. (20), respectively. These fitting curves are in good agreement with the experimental results.

It is noted that $V_0(\omega)$ is generated in conjunction with FMR, as shown in Fig. 6. The (red) solid line shows $V_0(\omega)$ induced by the in-plane RF field in the 22 nm-thick and 2 μm width wire located on the CPW, while the (black) dashed line shows the spectrum induced by the RF spin-polarized current in the 30 nm-thick wire of width 2.2 μm. The type of $V_0(\omega)$ due to the RF field is of the dispersion type, as described in Eq. (18), while that of $V_0(\omega)$ due to the RF spin-polarized current is of the Lorentzian resonant type. These results can be ascribed to the difference in torque direction; with the torque direction caused by the in-plane RF field perpendicular to the plane but that of the spin-transfer torque in plane.

## B. DC voltage induced by the out-of-plane RF field

The typical $V_0(\omega)$ spectra for $\theta=150^\circ$ for various $H_{\text{ext}}$ are shown in Fig. 7(a). Both the resonance frequency and $\Delta V_0$ increase with the magnetic field. Figure 7(c) shows $\Delta V_0(\omega)$ in $H_{ext}$=600 Oe for $\theta = 0^\circ \sim 180^\circ$. The resonance frequency takes a minimum at $\theta = 90^\circ$ and a maximum at $\theta = 0^\circ$ and $\theta = 180^\circ$, while $\Delta V_0(\omega)$ has a minimum at $\theta = 0^\circ, 90^\circ$ and $180^\circ$.

The additional and weak peaks observed in the trace of Figs. 7 (a) and 7(c) seem to be the additional spin wave modes. We note, however, only the dominant mode in order to proceed with the analytical calculation and the interpretation for the experimental data, since the additional modes are very hard to solve using the calculations. Figure 7(b) shows the static field dependence of the resonance frequency. The (red) solid circle is the experimental data, while the (black) solid line shows the fitting with the Kittel formula for the uniform precession mode given by Eq. (32) [18]. Effective fitting demonstrates how the magnetization vector of the wire is driven in the uniform precession mode with respect to the dominant mode. The inset of Fig. 7(b) shows the variation of $\Delta V_0(\omega)$ with magnetic field, which the (blue) open square is the experimental data. The $\Delta V_0(\omega)$ at the FMR frequency increases with increasing magnetic field. According to the result of Eq. (29), $\Delta V_0(\omega)$ at the FMR frequency $\omega_k$ is evaluated as follows:

$$\Delta V_0 \propto \omega_k^2 \Delta = \gamma_0^3 M_S H_c' H_b' \left( H_b' + H_c' \right). \tag{33}$$

When the demagnetization field is much larger than the applied field, we obtain $H_c' \approx M_S$ and $\Delta V_0 \propto H_b'$. These results (Fig. 7) indicate that the dominant mode is the uniform precessional mode, consistent with the model described by Costache [13].

The $\theta$ dependence of $\Delta V_0$ in (a) $H_{\mathrm{ext}}$=600 Oe and (b) $H_{\mathrm{ext}}$=50 Oe is shown in Figs. 8(a) and 8(b), respectively. The (red) solid circle is the experimental data. The (black) solid line in Fig. 8(a) shows a $-\sin 2\theta$ curve, which fits to the data well. This indicates that the uniform precession mode is dominated in the magnetic field to a sufficient extent to direct the magnetization parallel to the field. Conversely, the (black) solid line in Fig. 8(b) shows a $-\sin 2\theta \sin\theta$ curve, which fits the data relatively well. The angle dependence, however, is difficult to explain using solely the uniform precession mode generated by the RF field perpendicular to the plane. The origin of the $\theta$ dependence of $\Delta V_0(\omega_{\mathrm{k}})$ is contributable to the higher-order spin wave excitation within the confined structure of the shape-controlled ferromagnets [1 – 8].

In conclusion, we demonstrated the rectifying broadband FMR spectroscopy for the micron-scale ferromagnetic wires induced by the in-plane or out-of-plane components of RF field using the CPW method. We showed that the experimental DC voltage spectrum $V_0(\omega)$ produced in the wires can be well explained by the present phenomenological and analytical model; based on the coupling between the RF current and the AMR oscillation derived from the uniform magnetization precession.

We find that the characteristic asymmetric rectifying spectrum produced by the in-plane RF field differs in comparison with the rectifying effect due to the spin polarized current directly flowing through the wire, while the $\theta$ dependence of $\Delta V_0(\omega)$ is almost the same.

The broadband rectifying effect generated by the in-plane or out-of-plane components of the RF field using the CPW will provide a highly-sensitive measurement and analysis of spin dynamics in nano-scale magnets. As the effect can be easily tuned to control the resonance frequency, based on the shape of the wire or the magnetic field, this will facilitate fresh development for spintronics devices.

## Acknowledgements

The present work was partly supported by MEXT Grants-in-Aid for Scientific Research in Priority Areas, JSPS Grants-in-Aid for Scientific Research, and the Keio Leading-edge Laboratory of Science and Technology project 2007.

**Figure captions**

Figure 1

Schematic diagram and optical microscope image of the device in (a) in-plane and (b) out-of-plane magnetic field on the wire position. The external static magnetic field is applied in the substrate plane at an angle of $\theta$ from the longitudinal axis of the wire. A schematic diagram of electrical measurement circuit (c) and equivalent circuit (d) for the device illustrated in Fig. 1(a). (e) The schematic coordinate system adopted in this study.

Figure 2

The magnetic field distribution on the wire position on the center conductive strip line of CPW at the phase (a) $0^\circ$ and (b) $180^\circ$, calculated by HFSS. The magnetic field distribution on the wire position between the center strip line and the ground line of CPW at the phase (c) $0^\circ$ and (d) $180^\circ$. The phases $0^\circ$ and $180^\circ$ correspond to the direction of the current and are respectively the same and opposite as/to the positive $x$ direction, which is parallel to the longitudinal axis of the wire.

Figure 3

The impedance $Z(\omega)$ of the system comprising the wire on the center conductive strip line of the CPW as a function of the RF frequency $\omega/2\pi$.

Figure 4

(a) DC voltage spectra $V_0(\omega)$ generated by the 22 nm-thick $Ni_{81}Fe_{19}$ wire on the center strip line of the CPW in response to the RF current in the magnetic field of $|\boldsymbol{H}_{\mathrm{ext}}| = 100, 200, 400, 600$ and $800$ Oe for $\theta = 30^\circ$. (b) The frequency of the center position between the peak and the dip of the spectra as a function of the applied magnetic field. The line is a fit of Eq. (25) to the data. (c) $V_0(\omega)$ for different angles $\theta$ between the direction of the applied magnetic field and the longitudinal axis of the wire. Each resonant response is vertically shifted for clarity.

Figure 5

The (red) solid circles show the angle $\theta$ dependence of the DC voltage difference $\Delta V_0(\omega_{\mathrm{k}})$ at the FMR frequency in the magnetic field of (a) $|\boldsymbol{H}_{\mathrm{ext}}| = 900$ Oe and (b) $|\boldsymbol{H}_{\mathrm{ext}}| = 30$ Oe. The (black) solid lines represent the corresponding fits from case (i) and (ii) of Eq. (18) (see Text)

Figure 6

The rectifying spectra $V_0(\omega)$ generated by the in-plane RF field (the red solid line) and that by the RF spin-polarized current (the black dashed line) as a function of the RF frequency.

Figure 7

(a) DC voltage spectra $V_0(\omega)$ generated in the 30 nm-thick $Ni_{81}Fe_{19}$ wire prepared between the center strip line and the ground line of the CPW (Fig. 1(b)) in response to the RF current in the magnetic fields $|\boldsymbol{H}_{\text{ext}}| = 100, 200, 400, 600$ and 800 Oe for $\theta = 150^\circ$. (b) The frequency of the center position between the peak and the dip of the spectra as function of the magnetic field. The line is a fit of Eq. (25) to the data. $V_0(\omega)$ as a function of the magnetic field shown in the inset of Fig. 7(b). (c) $V_0(\omega)$ for different angles $\theta$ between the direction of the applied magnetic field and the longitudinal axis of the wire. Each resonant response is vertically shifted for clarity.

Figure 8

The (red) solid circles show the angle $\theta$ dependence of the DC voltage difference $\Delta V_0(\omega_k)$ at the FMR frequency in the magnetic field of (a) $|\boldsymbol{H}_{\text{ext}}| = 600$ Oe and (b) $|\boldsymbol{H}_{\text{ext}}| = 50$ Oe. The (black) solid line in Fig. 8(a) represents the corresponding fits from Eq. (29). The (black) solid line in Fig. 8(b) represents a $-\sin 2\theta \sin\theta$ curve. The fitting line is qualitatively in positive agreement with the data.

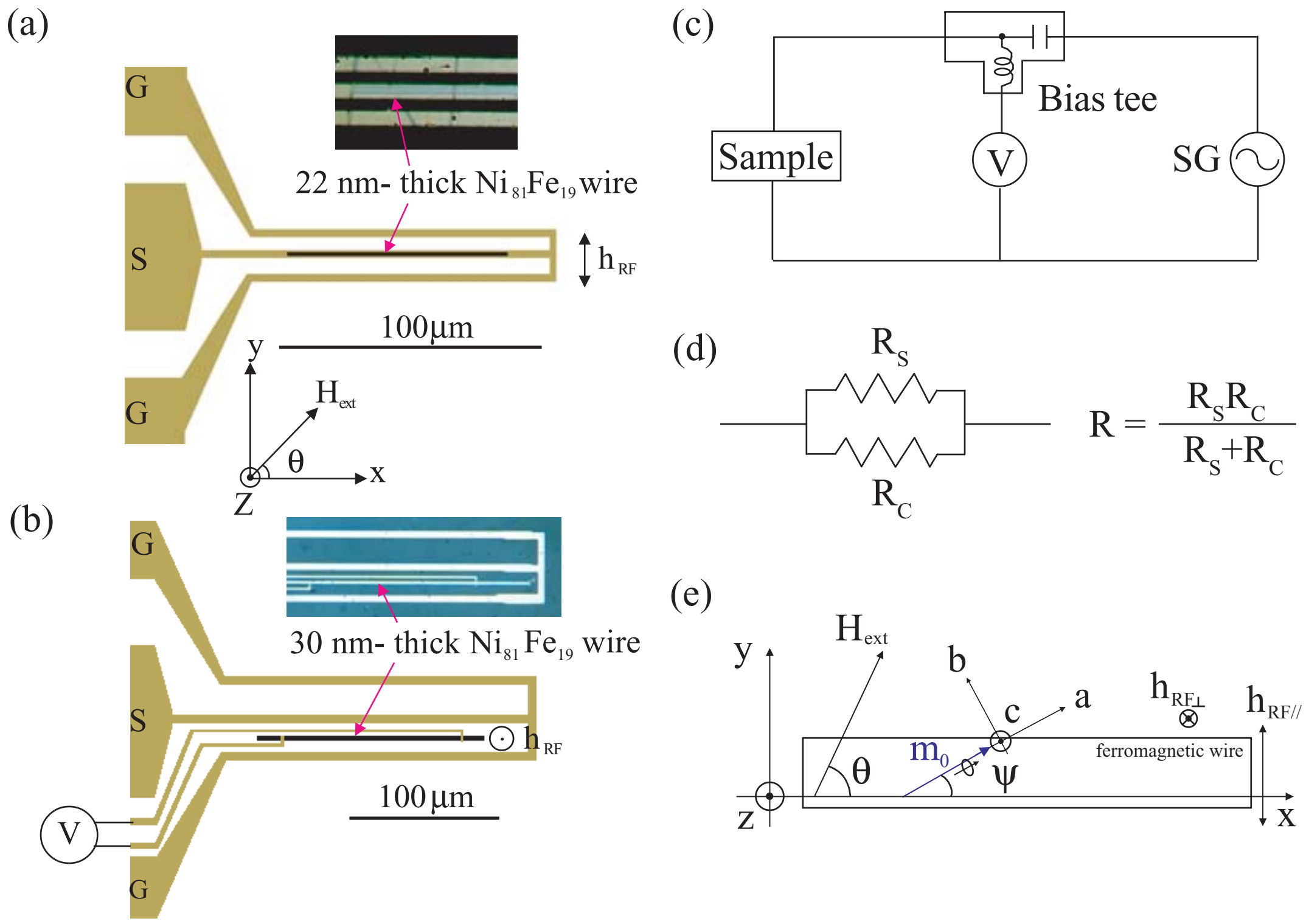

Fig. 1

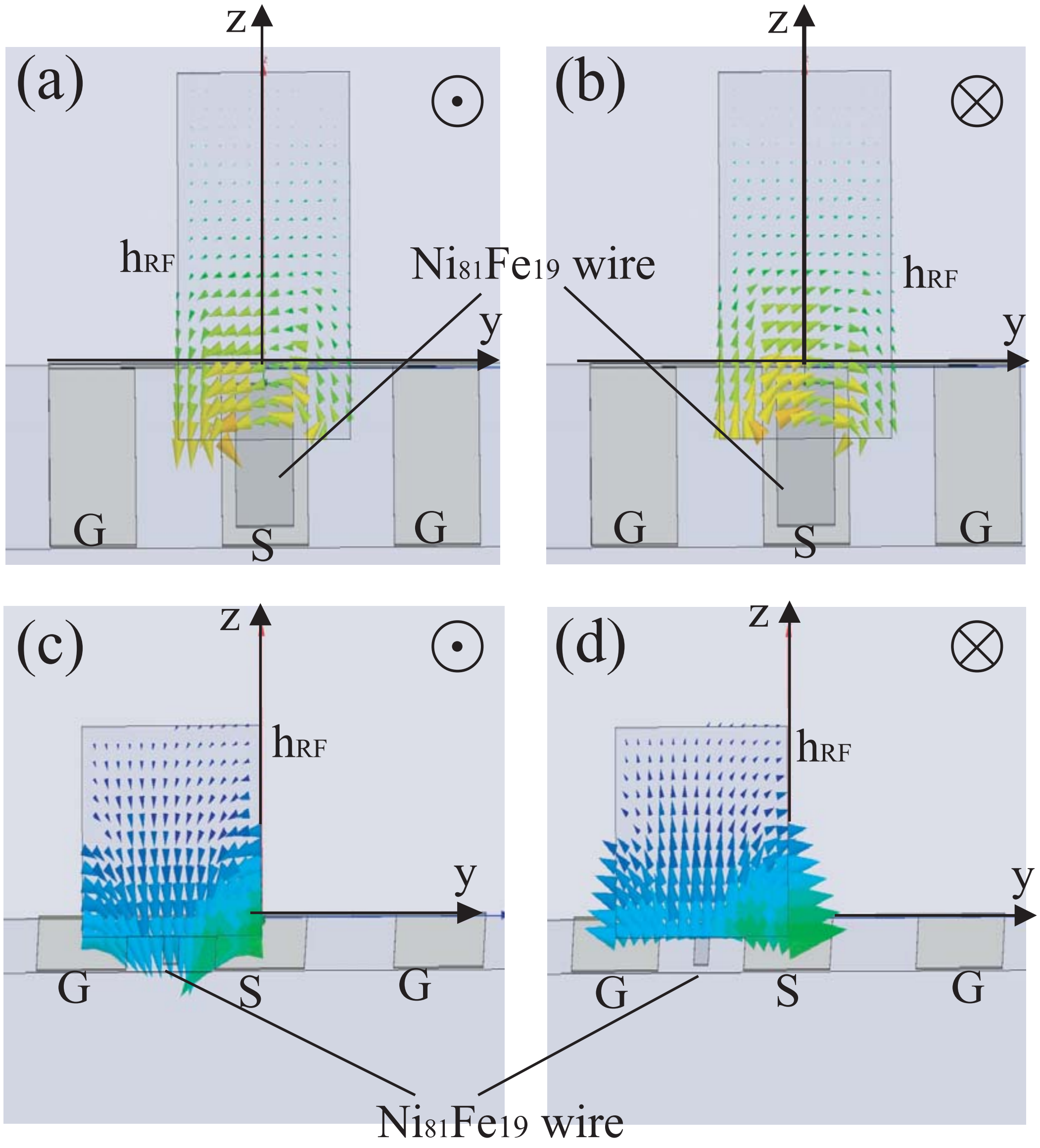

Fig. 2

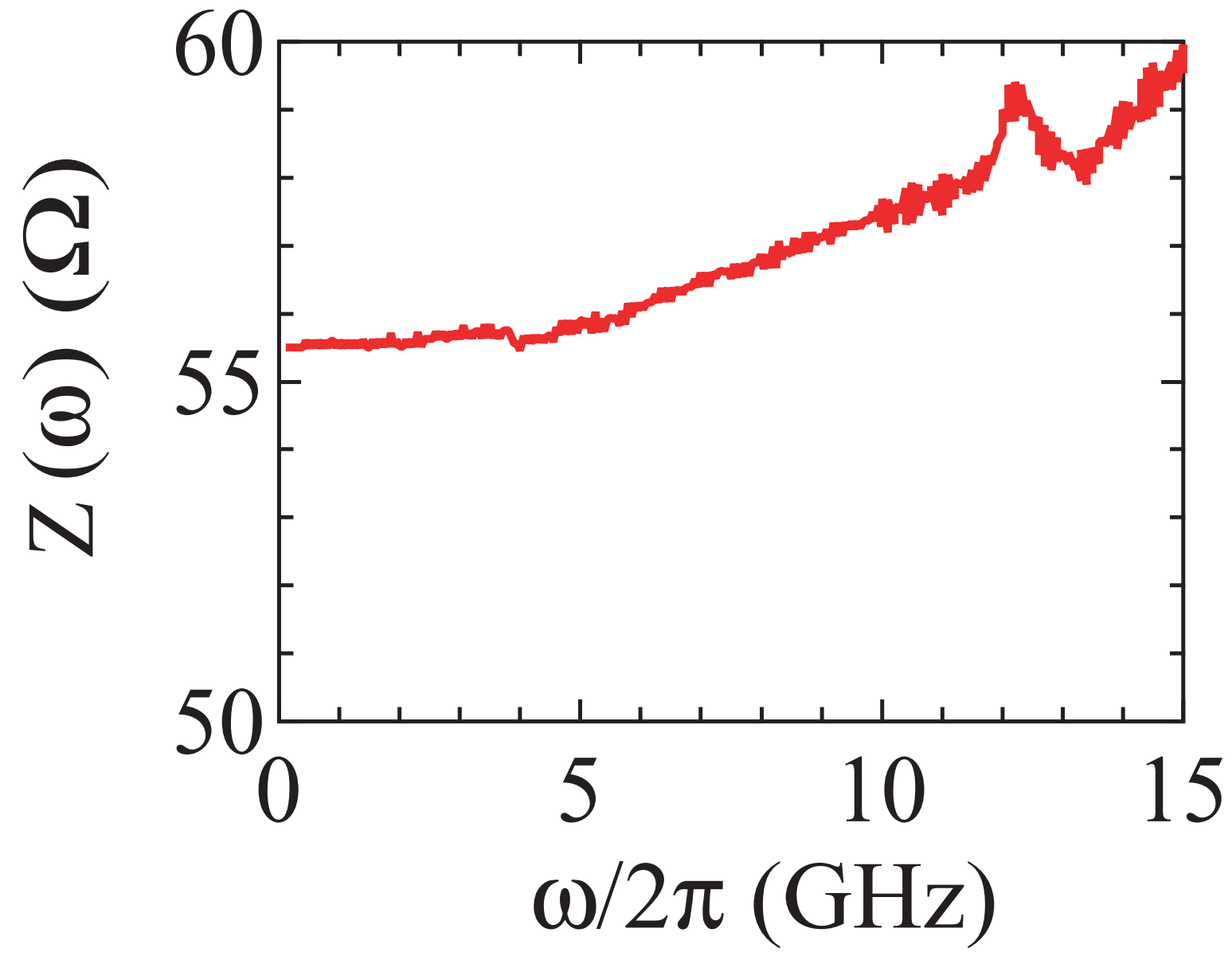

Fig. 3

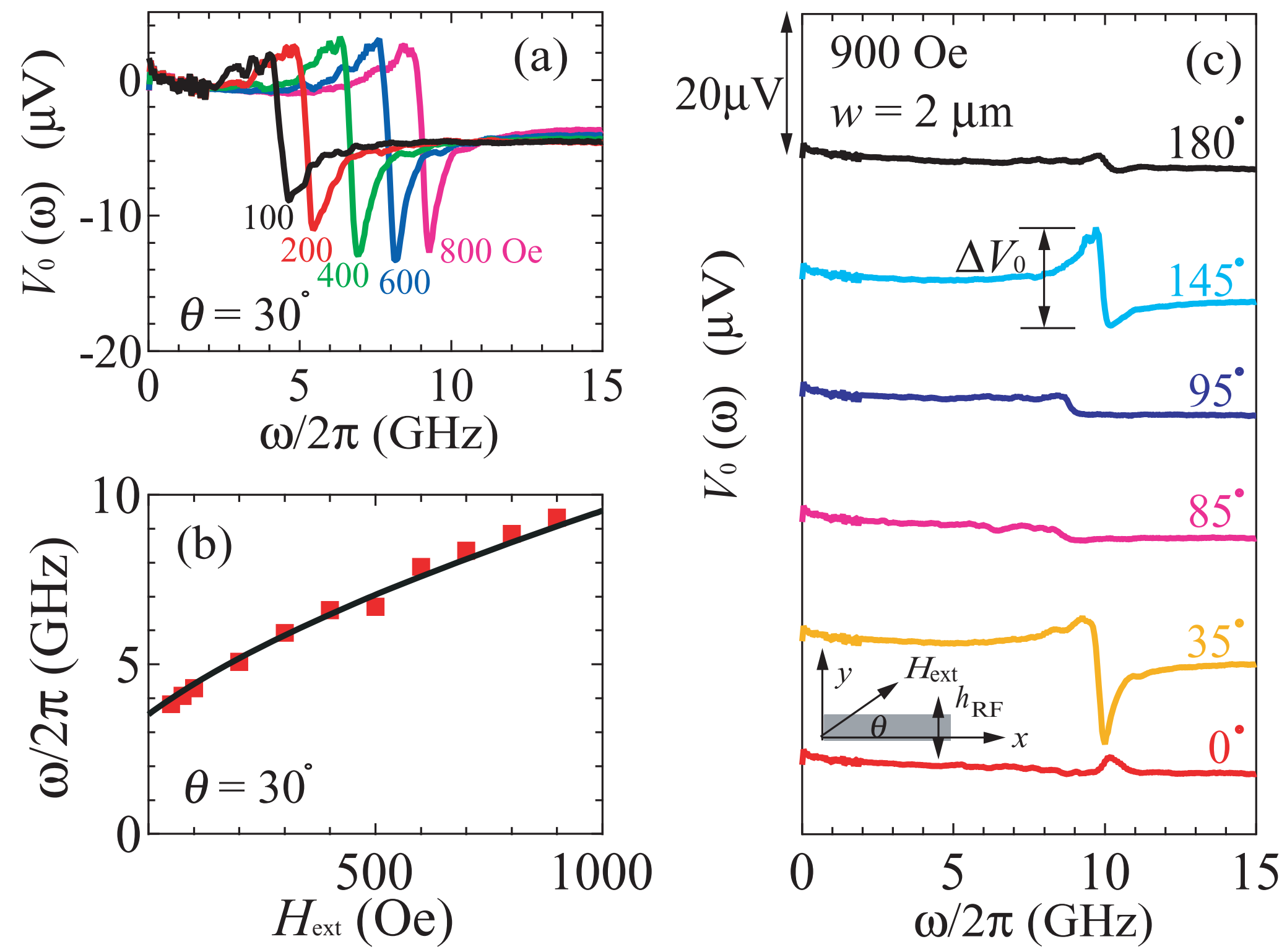

Fig. 4

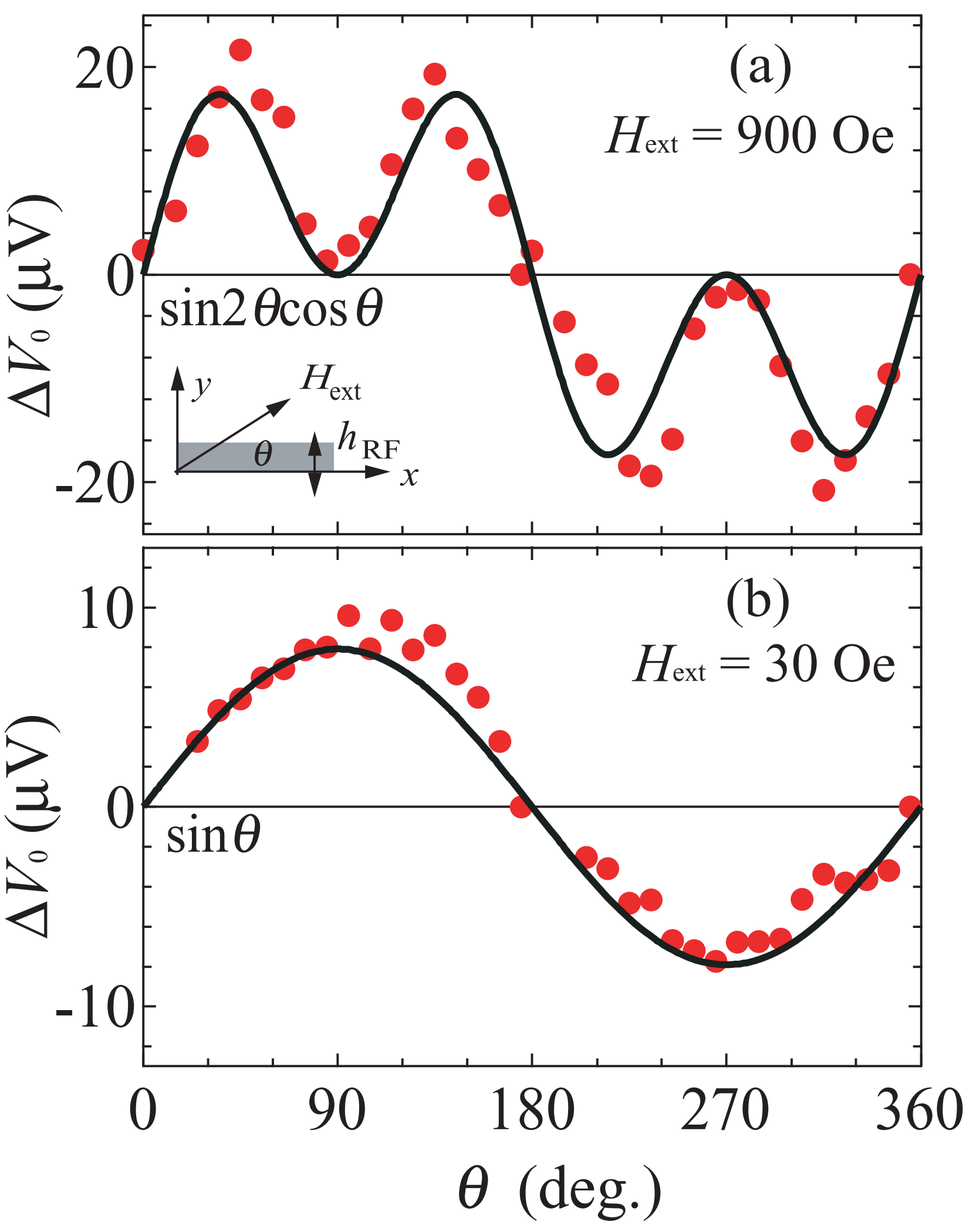

Fig. 5

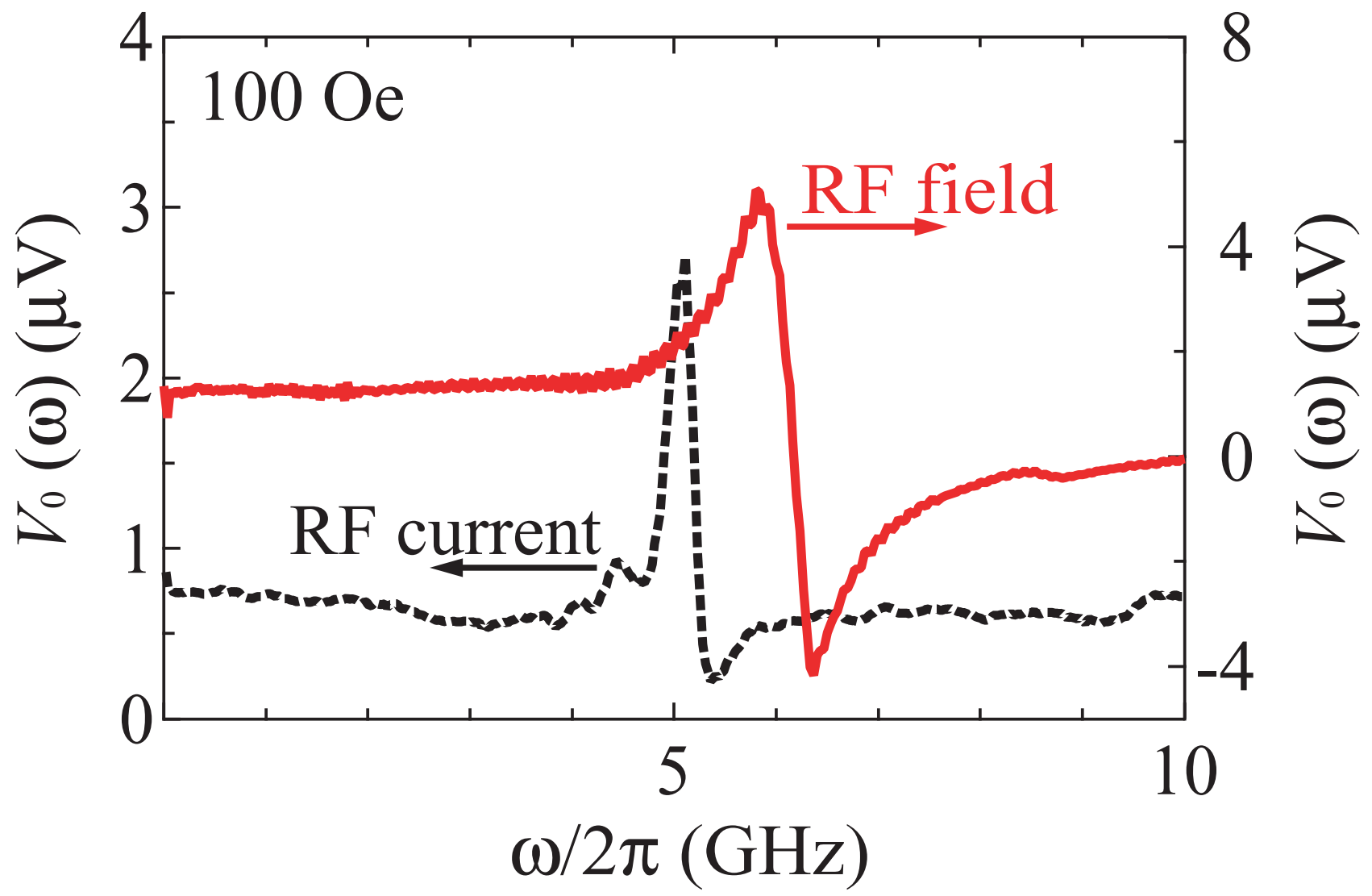

Fig. 6

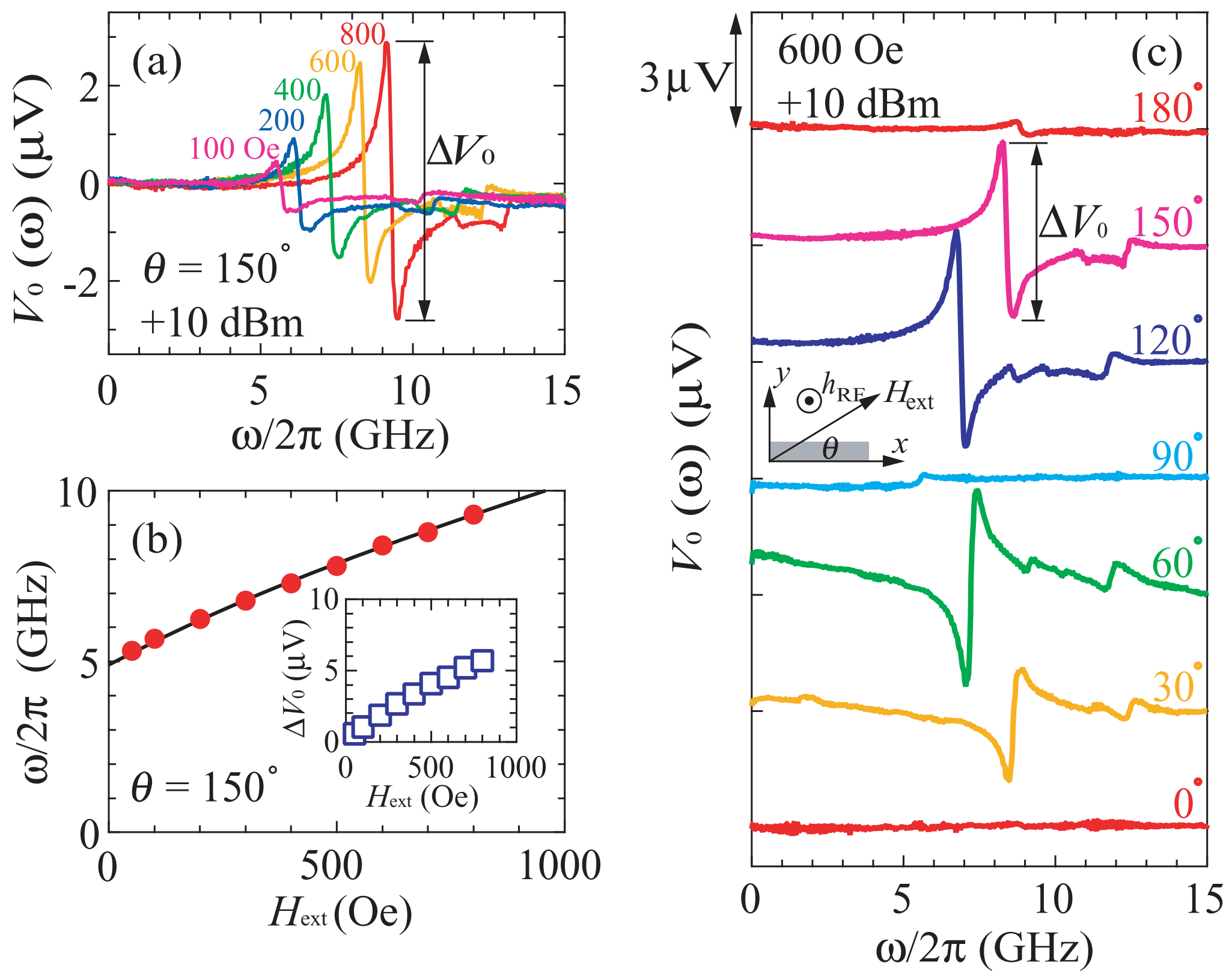

Fig. 7

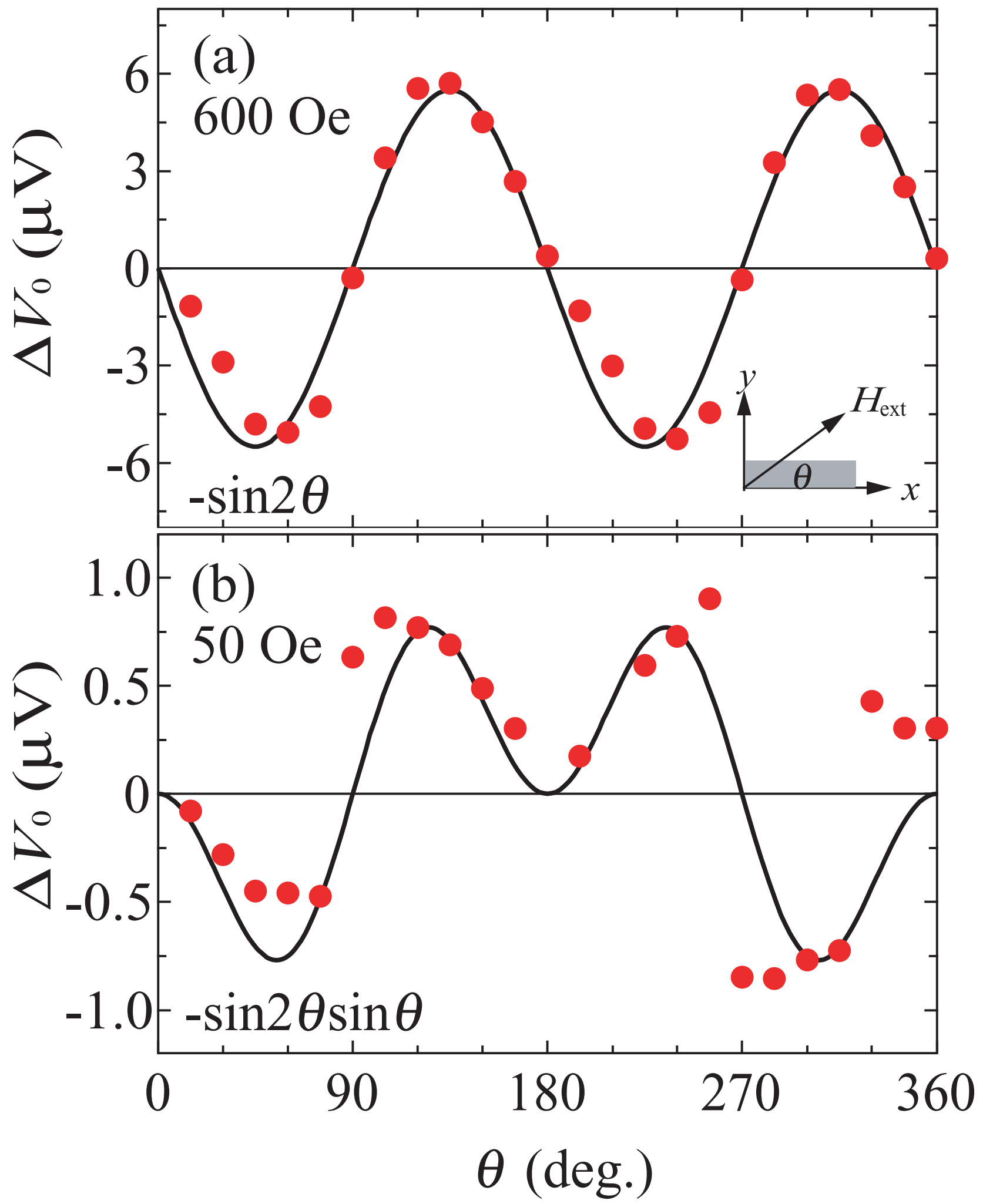

Fig. 8